\documentclass[pra,reprint,superscriptaddress,longbibliography]{revtex4-1}

\usepackage{amsmath}    
\usepackage{graphicx}   
\usepackage[lofdepth,lotdepth, caption=false]{subfig}
\usepackage{verbatim}   
\usepackage{color}      
\usepackage{hyperref}   
\usepackage{dcolumn}
\usepackage{amssymb}
\usepackage{xcolor}  
\usepackage{tabularx}

\begin{document}
\newcommand{\lcco}{La$_{2-x}$Ca$_{1+x}$Cu$_{2}$O$_{6}$}
\newcommand{\lccot}{La$_{1.9}$Ca$_{1.1}$Cu$_{2}$O$_{6}$}
\newcommand{\lccof}{La$_{1.85}$Ca$_{1.15}$Cu$_{2}$O$_{6}$}
\newcommand{\degrees}{$^\circ$C}
\newcommand{\Tc}{$T_{c}$}
\def\newr{\color{black}}
\def\newb{\color{black}}

\def\lno{La$_2$NiO$_4$}
\def\lnod{La$_2$NiO$_{4+\delta}$}
\def\lsno{La$_{2-x}$Sr$_x$NiO$_4$}
\def\lsco{La$_{2-x}$Sr$_x$CuO$_4$}
\def\lbco{La$_{2-x}$Ba$_x$CuO$_4$}
\def\lbcoate{La$_{1.875}$Ba$_{0.125}$CuO$_4$}
\def\lnsco{La$_{1.6-x}$Nd$_{0.4}$Sr$_x$CuO$_4$}
\def\lesco{La$_{1.8-x}$Eu$_{0.2}$Sr$_x$CuO$_4$}
\def\ybco{YBa$_2$Cu$_3$O$_{6+x}$}
\def\lcod{La$_2$CuO$_{4+\delta}$}

\title{Electron and hole contributions to normal-state transport\\ in the superconducting system Sn$_{1-x}$In$_x$Te}

\author{Cheng Zhang}
\thanks{Present address: Department of Materials Science and Engineering, University of Tennessee, Knoxville, Tennessee, USA}
\affiliation{Condensed Matter Physics and Materials Science Division, Brookhaven National Laboratory, Upton, New York 11973, USA}
\affiliation{Materials Science and Engineering Department, Stony Brook University, Stony Brook, New York 11794, USA}
\author{Xu-Gang He}
\affiliation{Condensed Matter Physics and Materials Science Division, Brookhaven National Laboratory, Upton, New York 11973, USA}
\affiliation{Department of Physics and Astronomy, Stony Brook University, Stony Brook, New York 11794, USA}
\author{Hang Chi}
\affiliation{Condensed Matter Physics and Materials Science Division, Brookhaven National Laboratory, 
Upton, New York 11973, USA}
\author{Ruidan Zhong}
\thanks{Present address: Department of Chemistry, Princeton University,
Princeton, New Jersey 08544, USA}
\affiliation{Condensed Matter Physics and Materials Science Division, Brookhaven National Laboratory, Upton, New York 11973, USA}
\affiliation{Materials Science and Engineering Department, Stony Brook University, Stony Brook, New York 11794, USA}
\author{Wei Ku}
\thanks{Present address: School of Physics and Astronomy, Shanghai Jiao Tong University, Shanghai 200240, China}
\author{Genda~Gu}
\author{J.~M.~Tranquada}
\author{Qiang Li}
\email{qiangli@bnl.gov}
\affiliation{Condensed Matter Physics and Materials Science Division, Brookhaven National Laboratory, Upton, New York 11973, USA}

\date{\today} 

\begin{abstract}

Indium-doped SnTe has been of interest because the system can exhibit both topological surface states and bulk superconductivity. While the enhancement of the superconducting transition temperature is established, the character of the electronic states induced by indium doping remains poorly understood. 
We report a study of magneto-transport in a series of Sn$_{1-x}$In$_x$Te single crystals with $0.1\le x \le 0.45$.  From measurements of the Hall effect, we find that the dominant carrier type changes from hole-like to electron-like at $x\sim0.25$; one would expect electron-like carriers if the In ions have a valence of $+3$. For single crystals with $x = 0.45$, corresponding to the highest superconducting transition temperature, pronounced Shubnikov-de Haas oscillations are observed in the normal state. 
In measurements of magnetoresistance, we find evidence for weak anti-localization (WAL). We attribute both the quantum oscillations and the WAL to bulk Dirac-like hole pockets, previously observed in photoemission studies, which coexist with the dominant electron-like carriers.
\end{abstract}

\maketitle
\section{introduction}

The discovery of topological insulators (TIs) has attracted great attention and stimulated considerable work on topological surface states arising from band inversion and time-reversal symmetry \cite{moor10,hasa10,qi11}. In topological states, electrons can flow with much reduced scattering from non-magnetic defects, offering great promise for next-generation electronics. Crystalline symmetry was soon identified as another promising route for obtaining the protected metallic surface states, leading to the new category of topological crystalline insulators (TCIs) \cite{fu11}. Tin telluride is a prototypical TCI predicted to have four conducting surface channels on specific crystallographic planes \cite{hsie12,xu12}. The band inversion has been confirmed, and surface states have been observed, by angle-resolved photoemission spectroscopy (ARPES) \cite{litt10,tana12}.  Experimental evidence for topologically non-trivial surface states has been obtained in transport studies of thin films \cite{assa14,akiy14}.  It has been proposed theoretically that combining topological surface states with bulk superconductivity may yield Majorana modes, which are of interest for use in quantum computing schemes \cite{fu08,qi09}.  Given that superconductivity can be induced in the SnTe system by indium doping, where the transition temperature can be as high as 4.5~K \cite{eric09,bala13,zhon13}, it is a natural system in which to look for the desired combination of states \cite{sasa12}.  

An unresolved issue concerns the nature of the carriers introduced by In doping.  Studies of IV-VI semiconductors have long indicated that In dopants act as if they contribute a resonance state or impurity band near the Fermi level \cite{kaid85}.  A relevant comparison is to Tl-doped PbTe, where the Tl$^{+1}$ and Tl$^{+3}$ states may be nearly degenerate \cite{mats06,gira18}.  Hall effect measurements on Sn$_{1-x}$In$_x$Te (SIT) with $x \lesssim0.1$ indicate than In induces an enhanced density of holes \cite{eric09,shen15}.  Angle-resolved photoemission spectroscopy (ARPES) studies of SIT have demonstrated the presence of small, hole-like Fermi pockets at the $L$ points of the Brillouin zone from both bulk and surface states for $x$ as large as 0.4 \cite{litt10,tana12,sato13,poll16}.  In contrast, recent measurements of the Hall effect on polycrystalline samples indicate a change in carrier type from holes to electrons on increasing $x$ beyond 10\% \cite{hald16}.  Indeed, supercell calculations of the band structure for SIT at small $x$ indicate the presence of an In-induced electron-like band crossing the Fermi level \cite{hald16}.

\begin{figure*}[t]
\begin{center}
\includegraphics[width=0.9\textwidth]{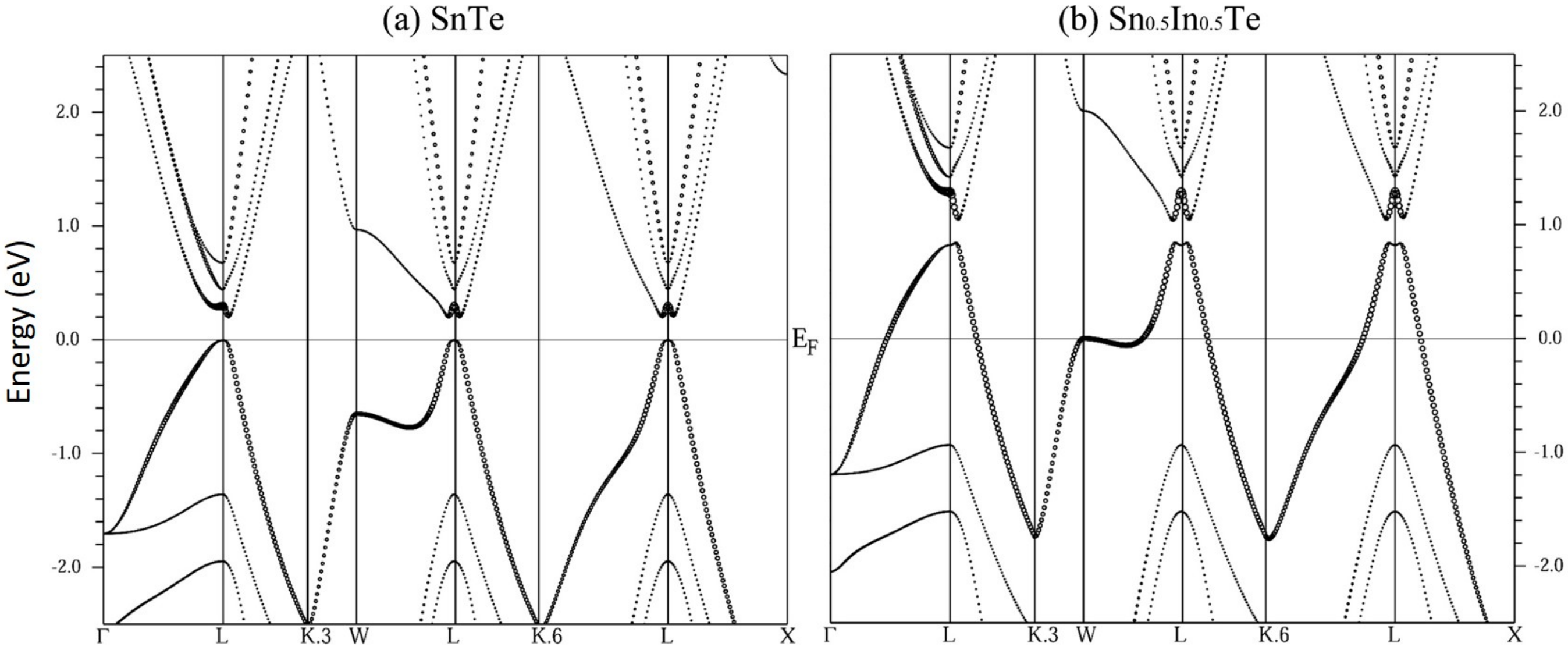}
\caption{First-principles band structure of (a) SnTe and (b) Sn$_{0.5}$In$_{0.5}$Te. Applying the VCA method to the occupancy of the Sn $5s$ orbital, the impact of the In substitution is to push the Fermi level down into the valence band,  reaching a level of 0.8~eV below the top of the valence band. The overall band structure remains intact, but with an enhancement of the band inversion.
\label{fg:bs}}
\end{center}
\end{figure*}

In this paper, we use transport measurements to explore the normal-state properties of Sn$_{1-x}$In$_x$Te single crystals for $0.1 \le x \le 0.45$, spanning most of the range of superconductivity.  From measurements of the Hall coefficient at $T = 5$~K, we infer the presence of both hole- and electron-like charge carriers, with a crossover in the dominant type at $x \sim 0.25$.  The significant change with increasing $x$ is the increase in electron mobility.  In field-dependent measurements of the Hall coefficient, we observe Shubnikov-de Haas oscillations, whose frequency and temperature-dependent amplitude are comparable to those expected for the bulk $L$-point hole pockets as detected by ARPES \cite{sato13,poll16}.  We also observe positive magnetoresistance that bears the signature of weak antilocalization (WAL).  As the magnitude of the magnetoresistance is independent of the orientation of the magnetic field, we attribute it to the bulk hole pockets and their Dirac-like character \cite{poll16,gara12}.  Overall, we find that the transport properties can be modeled in a consistent fashion by taking account of both the hole-like and electron-like carriers.

\section{Experimental Methods}

Single crystals of Sn$_{1-x}$In$_x$Te (SIT) with nominal In concentrations of $x = 0.10$--0.45 were grown by a modified floating-zone method. Pure SnTe used in the experiment was a polycrystalline sample, prepared via the horizontal unidirectional solidification method. The details were reported previously \cite{zhon13}. Crystals were cut into thin ($\sim 0.4$~mm) strips along (100) planes (with an orientational uncertainty of $5^\circ$), and measured in a Quantum Design Physical Property Measurement System (PPMS) equipped with a 9~T magnet.  A photo of a typical sample prepared for transport measurement is shown in the inset of Fig.~\ref{fg:trans}(a). The longitudinal resistivity was measured using a standard four probe method with in-line configuration. Hall measurement was conducted with voltage contacts placed on opposite sides of single crystals.  

\section{Band-structure calculations}

To provide context for interpreting the measurements, we did some simple band-structure calculations. We consider the case in which each indium dopant replaces a Sn atom and behaves as an acceptor, having one less electron than Sn. We used the {\tt WIEN2k} code \cite{blah90} to calculate the expected band structure using the virtual crystal approximation (VCA) to model the partial substitution of Sn by In. The results are shown in Fig.~\ref{fg:bs}. The main change due to 50\%\ In substitution is that the Fermi level moves deep into the valence band (0.8~eV from the top of the valence band), although the band inversion is also significantly enhanced compared to pure SnTe.  ARPES measurements on a film with $x = 0.41$ demonstrate that the Fermi level is indeed in the hole band \cite{poll16}, although the shift from $x = 0$ \cite{tana12} appears to be considerably smaller than the calculated value.

\section{Transport measurements}

\subsection{Doping dependence}

\begin{figure*}[t]
\begin{center}
\includegraphics[width=0.99\textwidth]{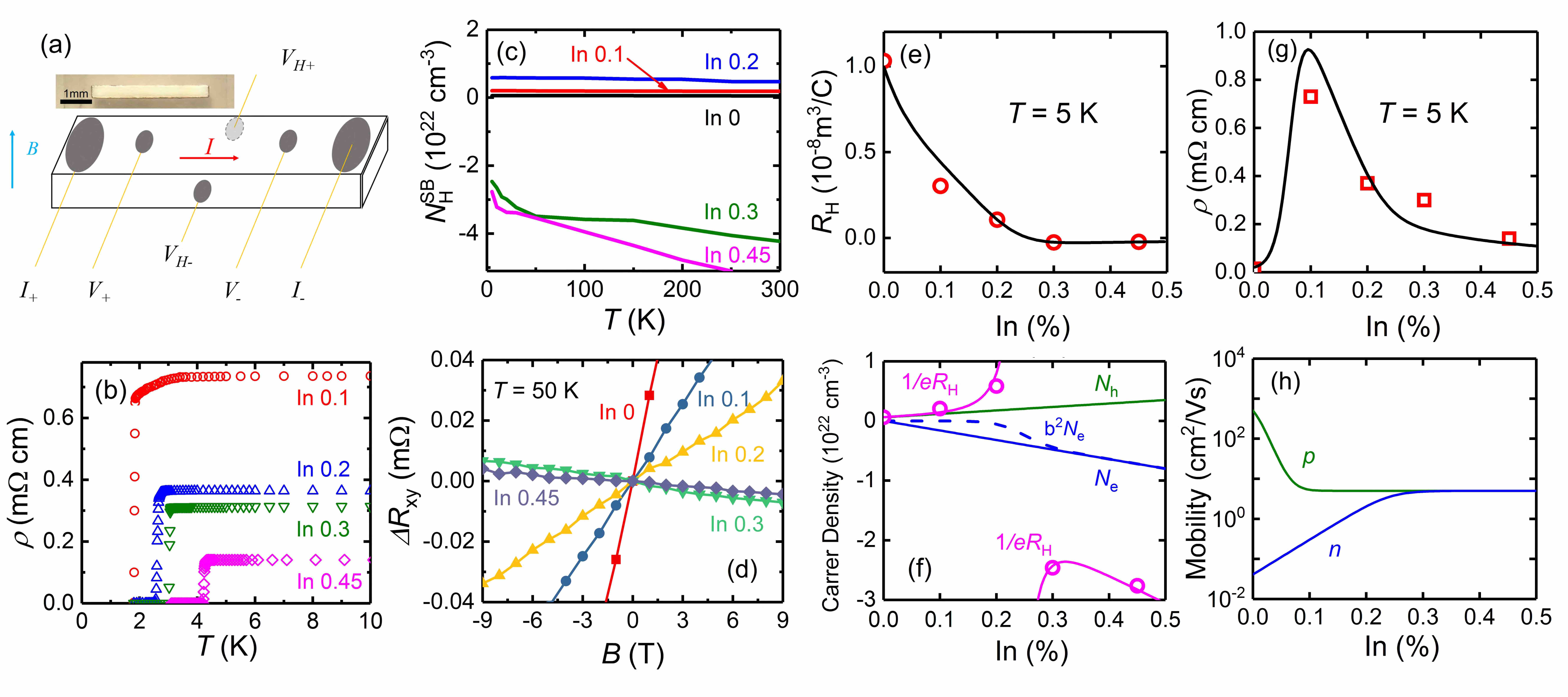} 
\caption{Transport measurements on Sn$_{1-x}$In$_x$Te samples.  {\newr (a) Sketch of the contact locations on the sample. Inset shows a typical single crystal sample prepared for transport measurements (length $\sim6$~mm).  (b)} Resistivity vs.\ temperature for various In concentrations; note that the superconducting transition increases with $x$.  {\newr (c)} Temperature dependent Hall carrier concentration $N_{\rm H}$ calculated using single band model $N_{\rm H} = 1/eR_{\rm H}$. A change of dominant carrier type occurs between $x = 0.2$ and 0.3. {\newr (d) Transverse resistance $R_{xy}$ as a function of magnetic field $B$, at $T=50$~K.  (e)} Hall coefficient at 5~K vs.\ In doping (circles); line is a fit with the two-band model described in the text. {\newr (f)} Plot of carrier concentrations $N_h$ (green line) and $N_e$ (blue line) assumed in the model calculation; dashed line represents the $N_e$ mutiplied by the squared ratio of mobilities, as discussed in the text.  Circles indicate $1/eR_{\rm H}$ data; magenta line is the model calculation.
{\newr (g)} Resistivity at 5~K (squares), compared with the model calculation (line). (Data point at $x=0$ from \cite{eric09}.)  {\newr (h)} Hole and electron mobilities used in the model calculations.
\label{fg:trans}}
\end{center}
\end{figure*}

The transport data for our SIT crystals spanning a range of In concentrations are presented in Fig.~\ref{fg:trans}. In particular, longitudinal electrical resistivity is shown as a function of temperature in Fig.~\ref{fg:trans}(b), where the superconducting transition temperature clearly increases while the magnitude of the resistivity decreases with In-doping. Figure~\ref{fg:trans}(c) shows the temperature dependence of the carrier concentration $N_{\rm H}$ calculated from the Hall coefficient $R_{\rm H}$ using a single band model: $N_{\rm H} = 1/(eR_{\rm H})$ (positive values for holes and negative for electrons), where $e$ is the electron charge. A dramatic change in the carrier type from $p$ type to $n$ type is found between $x = 0.2$ and 0.3. The sign change is qualitatively consistent with the results of Haldolaarachchige {\it et al.} \cite{hald16} measured on polycrystalline samples.  
{\newb (We do not have a clear understanding of the quantitative difference with \cite{hald16} on the In concentration at which the single-band $N_{\rm H}$ changes sign.  While there could be a small difference in concentration of Sn vacancies, that is unlikely to be the explanation, as the measured values of lattice parameter and superconducting transition temperature as a function of $x$ \cite{hald16,zhon13} are rather consistent, which would not be the case for a significant difference in Sn vacancies \cite{bush86}.)}

The apparent sharp jump in carrier concentration and sign with doping {\newr is surprising.}
If we look at the measured Hall coefficient at a temperature of 5~K, shown by the circles in Fig.~\ref{fg:trans}(e), we see that it varies smoothly with doping.  {\newr To understand what may be going on, we consider the behavior of the In dopants.  An In atom has an outer $5s^25p^1$ configuration.  When doped into SnTe, it will certainly give up its outer $5p$ electron to yield In$^{+1}$.  Past work \cite{kaid85} has demonstrated that the In $5s$ level lies below the chemical potential of SnTe.  For reference, the In-Te bond length in an In$^{3+}$ compound such as In$_2$Te$_3$ is 2.67~\AA\ \cite{wool59}, whereas the Sn-Te bond length in SnTe is 3.16 \AA\ \cite{zhon13}; hence, it is plausible that the $5s$ electrons of an isolated In dopant will not hybridize with Te neighbors.  Hybridization between In ions will only occur as the probability of In dopants being near one another becomes significant.  Indeed, band structure calculations by Haldolaarachchige {\it et al.} \cite{hald16} for $x=0.12$ yield a narrow extra band, largely below the chemical potential.  If the chemical potential shifts into this band, then the dominant carriers may become electron-like.} 

\begin{figure}[b]
\begin{center}
\includegraphics[width=\columnwidth]{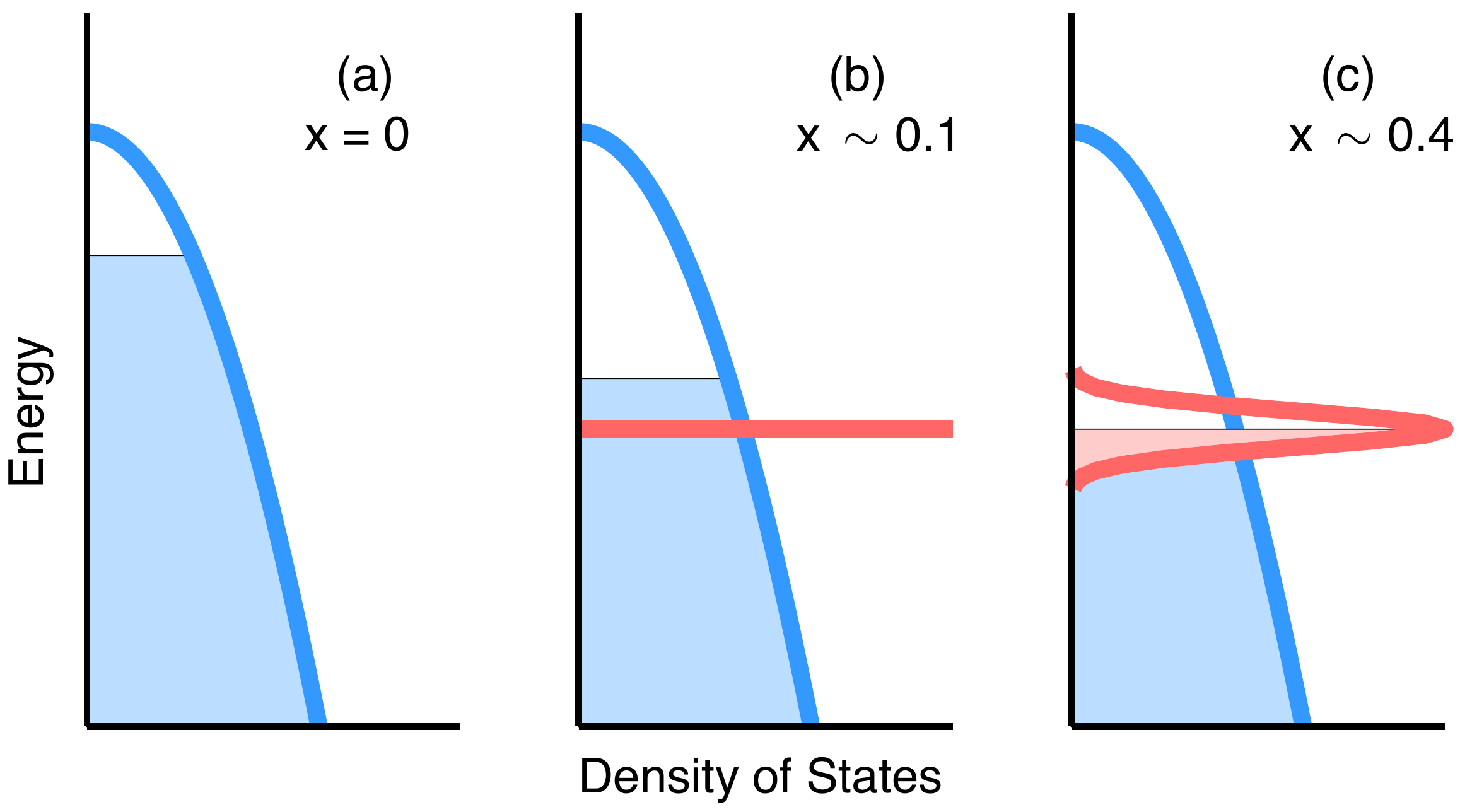} 
\caption{\newr Cartoon of the hole (blue) and electron (red) densities of states vs.\ energy for Sn$_{1-x}$In$_x$Te, as discussed in the text.  (a) $x=0$; (b) $x\sim0.1$; (c) $x\sim0.4$.
\label{fg:dos}}
\end{center}
\end{figure}

\begin{figure*}[t]
\begin{center}
\includegraphics[width=0.65\textwidth]{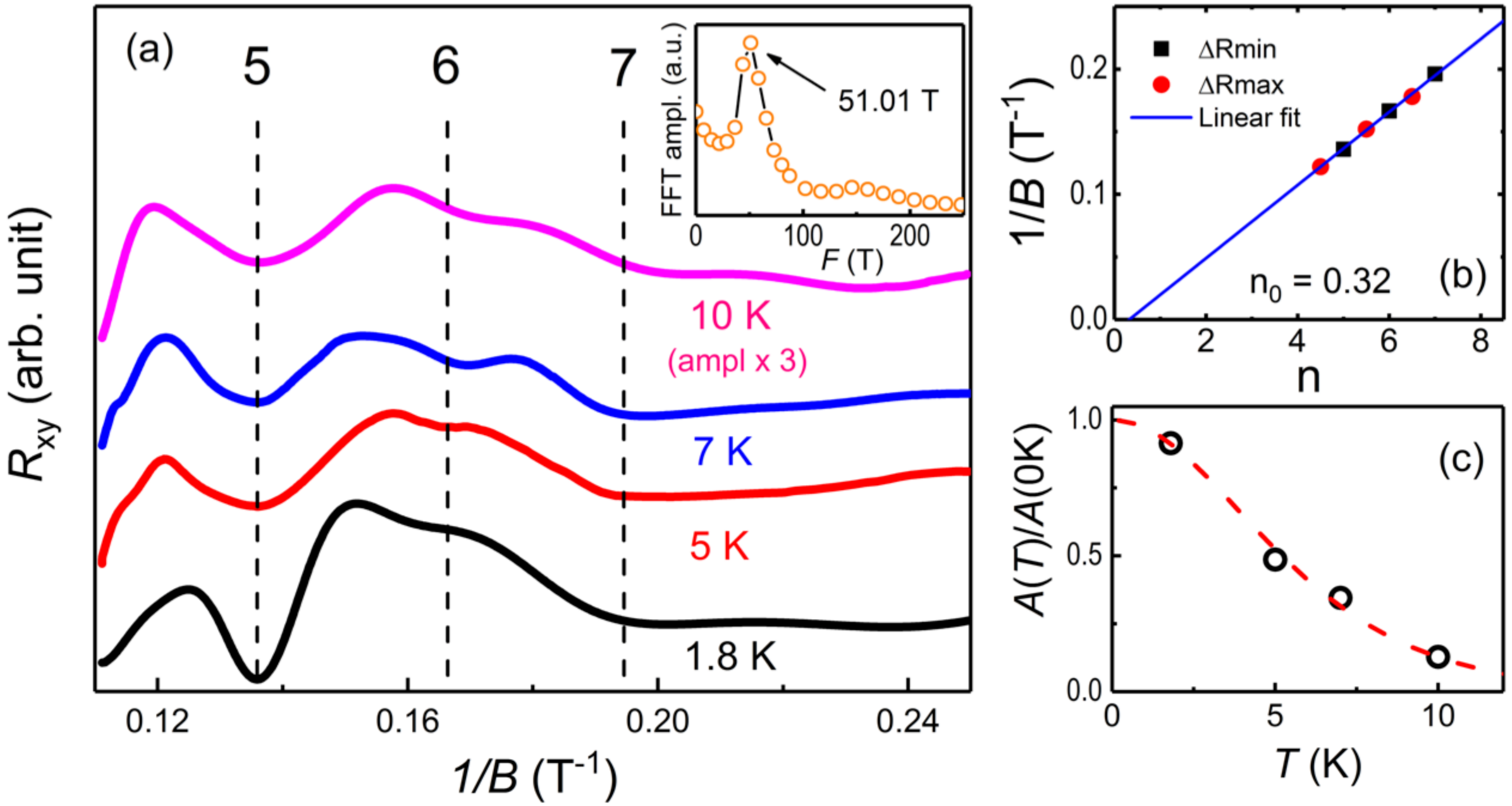} 
\caption{(a) SdH oscillations in Sn$_{0.55}$In$_{0.45}$Te transverse resistance measured at temperatures of 1.5 to 10~K plotted vs.\ inverse magnetic field, after subtraction of conventional Hall response.  The 10-K data are multiplied by 3, and curves have been offset vertically. The assigned Landau level indices are indicated by the numbered vertical dashed lines.  Inset shows Fast Fourier transform (FFT) spectrum of the 5~K data.  (b) Plot of inverse field for oscillation extrema vs.\ Landau level index; linear fit yields an intercept of $0.32 \pm 0.07$. (c) The temperature dependence of the oscillation amplitude at $n = 5$ fitted by Lifshitz-Kosevich theory (dashed line), yielding a cyclotron mass $m_{\rm cycl}$ of $0.185 m_e$.
\label{fg:qo}}
\end{center}
\end{figure*}

{\newr We imagine a scenario as illustrated in Fig.~\ref{fg:dos}{\newb; note that our synthesis differs in some details from previous work \cite{kaid85,hald16}}.  For $x=0$, the chemical potential of SnTe lies in the valence band, due to a small density of Sn vacancies.  For small but finite $x$, the In $5s$ electrons are localized near the In dopant sites, with the energy level lying below the initial chemical potential.  Because of the $5s$ localization, the In ions act effectively as In$^{+1}$, causing the chemical potential to drop.} This is consistent with previous transport results \cite{eric09,bush84,shen15} that $N_h$ is finite even at $x=0$, grows with $x$ up to at least $x\sim0.1$, while ARPES studies \cite{sato13,poll16} suggest that the hole pockets continue to grow slowly at larger $x$. {\newr At large enough $x$, the In $5s$ states form a narrow band and the chemical potential gets pinned in this band.  At this point, the In ions act as In$^{3+}$ and electron-like carriers become important.  }

{\newr To approximately describe this behavior, we consider a two-band} model that contains contributions from both holes and electrons \cite{ashc76}:
\begin{equation}
  R_{\rm H} = {(N_h - N_e b^2) \over e(N_h + N_e b)^2},
  \label{eq:RH}
\end{equation}
where $N_h$ ($N_e$) is the density of holes (electrons) and $b = \mu_e / \mu_h$, the ratio of mobilities of the electrons and holes.  {\newr We take $N_h$ to be} small but finite at $x = 0$ and {\newr allow it to grow} linearly with $x$.  In contrast, we take $N_e$ to be equal to the density of In ions, {\newr but make the mobility $\mu_e$ very small in the regime where the electrons are localized.} The key to the crossover in dominant carrier type is the {\newr variation in the} mobility ratio, $b$, {\newr which} 
starts out small, but then grows rapidly towards one at larger $x$; the product $N_e b^2$ is indicated by the dashed line in Fig.~\ref{fg:trans}(f).  With these choices, we obtain the solid line in Fig.~\ref{fg:trans}(e), which gives a good description of $R_{\rm H}(x)$. 

Of course, in modeling $R_{\rm H}$ \footnote{{\newr We note that there are field-dependent corrections to Eq.~(\ref{eq:RH}) which predict curvature in $R_{xy}$ that are not seen in the data of Fig.~\ref{fg:trans}(d).  We simply take this as a shortcoming of our crude model.} } we have introduced more degrees of freedom than we have constraints.  To test the model further, it is useful to consider the magnitude of $\rho$, which depends on both the carrier densities and the mobilities.  {\newr The mobility can be quite large for SnTe at low temperature \cite{allg72}, but even 1\%\ In doping raises the resistivity almost two orders of magnitude \cite{eric09}, implying a huge drop in mobility.  From the reported resistivity for $x=0$ \cite{eric09} and our measurement of $R_{\rm H}$, we estimate an initial hole mobility of }  $\sim500$~cm$^2$/(V s) {\newr (small compared to values in \cite{allg72}); it} then drops rapidly on the introduction of In, decreasing by two orders of magnitude by $x\sim0.1$.  We assume that the hole mobility then remains constant at 5~cm$^2$/(V s).  Meanwhile, the electron mobility starts out at a negligible level {\newr (where the electrons are localized)} and steadily rises, becoming comparable to the hole mobility at $x\sim0.25$.  Using the model mobilities plotted in Fig.~\ref{fg:trans}(h) together with the carrier densities shown in Fig.~\ref{fg:trans}(f), we obtain for $\rho$ the solid line plotted in Fig.~\ref{fg:trans}(g), which certainly captures the trend of the experimental data points.

\subsection{Quantum Oscillations}

Measurements of the field dependence of $R_H$ for the $x=0.45$ sample {\newr with the field along (001)} revealed prominent SdH oscillations. Figure~\ref{fg:qo}(a) shows the oscillations in the transverse resistance at 10~K and below, after subtracting backgrounds, revealing periodic behavior as a function of inverse field. The positions of the peaks and valleys appear to be independent of temperature, though the magnitude is not.  Analysis of these features can provide parameters related to the relevant portions of the Fermi surface. The inset in Fig.~\ref{fg:qo}(a) shows the amplitude of the Fourier transform of the 5-K SdH spectrumm yielding the frequency $f_{\rm SdH} = 51$~T.  The cross section of the Fermi surface, $A_{\rm F}$ is related to the SdH oscillation frequency via the Onsager relation \cite{bao12}: $f_{\rm SdH}=(h/4\pi^2e)A_{\rm F}$, where $A_{\rm F}=\pi k_{\rm F}^2$, with $k_{\rm F}$ being the Fermi wave vector.   The resulting $k_{\rm F}$ is 0.04~\AA$^{-1}$.
 
\begin{figure*}[t]
\begin{center}
\includegraphics[width=0.8\textwidth]{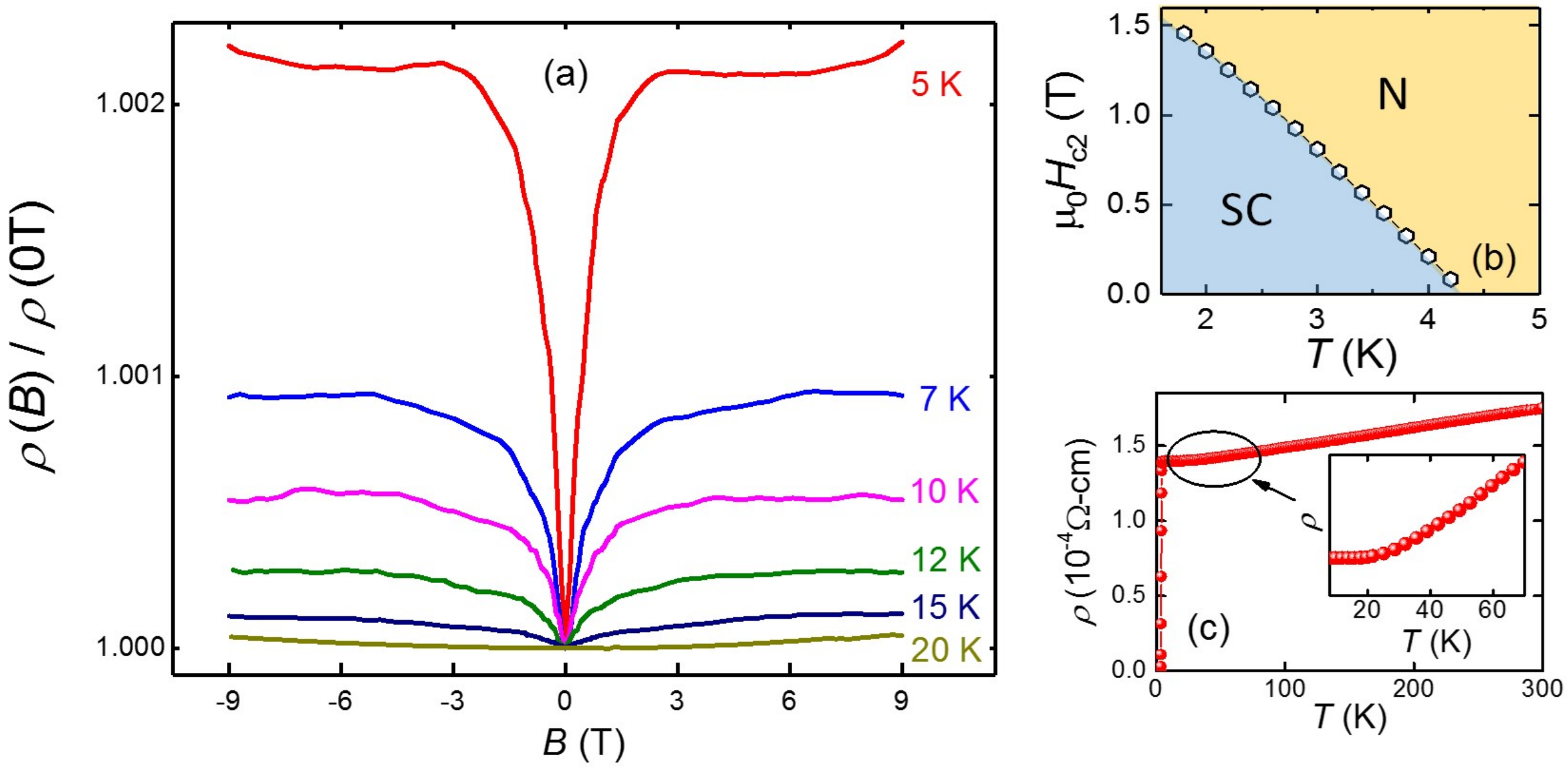} 
\caption{Normalized longitudinal resistivity for the Sn$_{0.55}$In$_{0.45}$Te single crystal. (a) Magnetoresistance curves exhibit a sharp cusp at $B=0$ with an amplitude that diminishes with increasing temperature. (b) Phase diagram of Sn$_{0.55}$In$_{0.45}$Te showing the upper critical field vs.\ temperature \cite{zhon13}. (c) Zero-field resistivity over an extended temperature range; inset shows that there is a plateau below 20~K, where the MR develops.
\label{fg:wal}}
\end{center}
\end{figure*}

 The Landau level index has been assigned as done in previous reports \cite{qu10,bao12,xion12,ren10}, and the positions of the peaks and valleys measured in inverse field are plotted as a function of Landau level index $n$ in Fig.~\ref{fg:qo}(b). The linear fit leads to a non-zero intercept of $0.32 \pm 0.07$, a value comparable to 0.5 which is expected for massless Dirac fermions, more commonly for surface states in TIs \cite{qu10,bao12,xion12,ren10,anal10,hsiu13}.
 
{\newr Given the substantial carrier density in our sample, we expect that the quantum oscillations must come from bulk states.  Of course, we have both hole- and electron-like carriers, so which of these contributes the oscillations?  ARPES studies of SIT have observed the hole-pockets near the $L$ points, and have distinguished bulk and surface states by their dispersion with momentum perpendicular to the sample surface \cite{sato13,poll16}.  For a sample with $x\approx0.4$, both the bulk and surface states show a Dirac-like dispersion near the hole pockets, while no electron-like features have been identified \cite{poll16}.  Hence, it seems most plausible to associate the oscillations with bulk hole-like pockets.}
 
 Figure~\ref{fg:qo}(c) shows the temperature dependence of the SdH amplitude $A(T)$ at $n = 5$, fitted with the Lifshitz-Kosevich theory \cite{shoe84}: $A(T)=\lambda/\sinh \lambda$, where $\lambda = (\pi k_{\rm B}T/ehB)m_{\rm cycl}$. The cyclotron mass $m_{\rm cycl}$ is found to be $0.185 m_e$ at a field of 6.5~T, where $m_e$ is the free electron mass.  Assuming a Dirac-like dispersion, the Fermi velocity $v_{\rm F}$ can be calculated by $v_{\rm F}m_{\rm cycl} = \hbar k_{\rm F}$ \cite{bao12,ren10}, yielding $2.5 \times 10^5$ m/s. 
 
We can compare our results with those obtained by ARPES for the bulk $L$-point pockets of a (111) SIT film with $x\approx 0.4$ \cite{poll16}.  The latter study found a linear dispersion characterized by $k_{\rm F} =0.095$~\AA$^{-1}$ and a Fermi velocity of $6.0\times 10^5$~m/s, which puts the Fermi level 0.38~eV below an extrapolated Dirac point.  This compares with our $k_{\rm F} = 0.04$~\AA$^{-1}$ and $v_{\rm F} = 2.5\times 10^5$ m/s, which would put the Fermi level at 0.07 eV.  The main point here is that the values are of {\newr comparable} magnitude.
 
 \subsection{Magnetoresistance}

{\newr Observations of WAL \cite{hika80} in low-temperature magnetoresistance measurements on pristine SnTe have been used to identify the presence of topologically-protected surface states \cite{akiy14,assa14}.  Recent theoretical work has demonstrated that one can also observe WAL from bulk Dirac-like states with strong spin-orbit coupling \cite{gara12}.  Now, WAL from surface states should be sensitive to the orientation of the magnetic field with respect to the surface \cite{he11}.  Below we demonstrate WAL that is insensitive to field direction, consistent with bulk Dirac-like states, which, based on the analysis of the SdH oscillations, are likely associated with the hole-like pockets near the $L$ points.}

Figure~\ref{fg:wal}(a) shows the normalized {\newr longitudinal magnetoresistance (MR)} $\rho(B)/\rho(0~{\rm T})$ obtained with the magnetic field applied perpendicular to the current {\newr for the Sn$_{0.55}$In$_{0.45}$Te sample}. The magnitude of the MR increases rapidly on cooling below 20 K.  Note that we are limited in temperature range by the superconducting transition; for reference, the superconducting phase diagram is shown in Fig.~\ref{fg:wal}(b).  The rapid rise and saturation looks very much like the WAL that has been observed in association with topologically-protected surface states in TIs such as Bi$_2$Te$_3$ \cite{he11}.  Similar behavior was observed for our $x = 0.3$ sample, but with a reduced magnitude.  Figure~\ref{fg:wal}(c) shows the temperature dependence of the zero-field resistivity, indicating a saturation below 20~K, corresponding to the region where the sharp MR appears.

\begin{figure}[t]
\begin{center}
\includegraphics[width=0.45\textwidth]{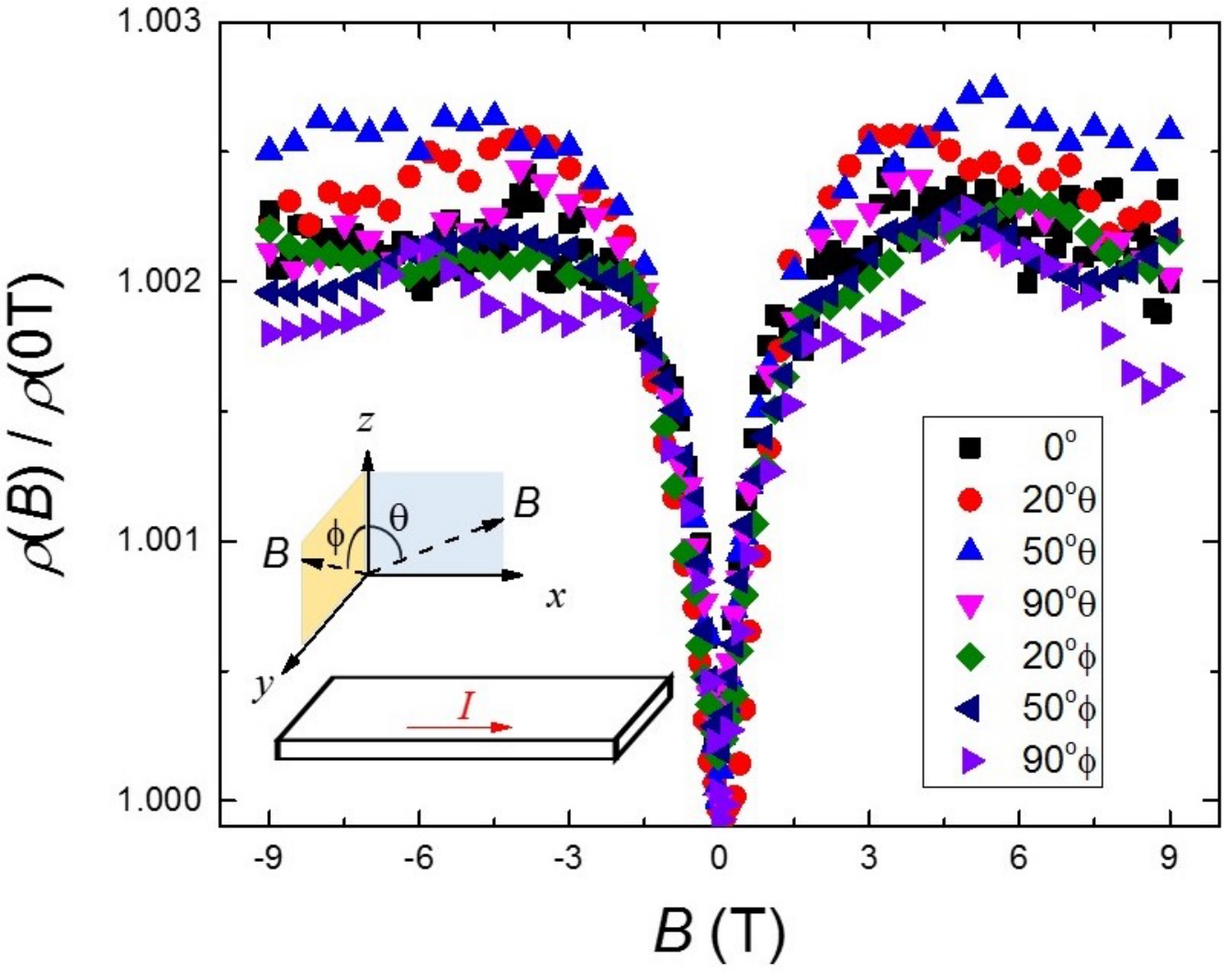} 
\caption{Angle-dependent magnetoresistance measurements Sn$_{0.55}$In$_{0.45}$Te at 5~K. The cusp appears to be independent of orientation of applied magnetic field. Inset defines the orientation angles of the applied magnetic field relative to the sample surface and direction of applied current.
\label{fg:angle}}
\end{center}
\end{figure}
 
{\newr As noted above,} MR from surface states should be sensitive to the orientation of the magnetic field.  To test this, angle-dependent MR was measured at 5~K. As shown in the inset of Fig.~\ref{fg:angle}, $\theta$ and $\phi$ denote the angles between the magnetic field and $z$ axis within $x$-$z$ and $y$-$z$ planes, respectively, where the electrical current is always applied along the $x$ direction. We observe that the low-field MR is essentially independent of angle.  This isotropic response indicates bulk behavior.

{\newr For WAL from bulk states \cite{gara12},}  the contribution to the conductance has the same form as that for the two-dimensional case \cite{hika80}:
\begin{equation}
  \Delta G = \alpha {e^2\over \pi h} \left[\ln\left({B_\phi\over B}\right) - \psi\left({B_\phi\over B}+\frac12 \right)\right],
  \label{eq:G}
\end{equation}
where $\psi$ is the digamma function and $B_\phi=\phi_0/(8\pi l_\phi^2)$, with $\phi_0 = h/e$ and $l_\phi$ being the phase coherence length.  The parameter $\alpha$ is a constant that equals 1 for the case of Dirac-like dispersion in a single pocket at the Brillouin zone center \cite{gara12} (which is slightly different from our case of pockets at the $L$ points).

\begin{figure}[t]
\begin{center}
\includegraphics[width=0.45\textwidth]{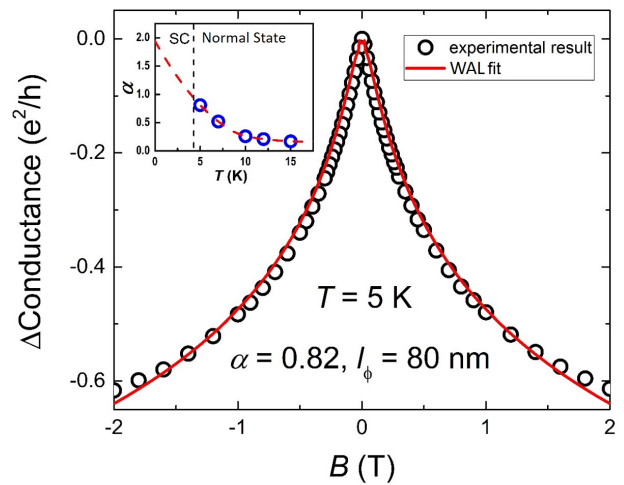} 
\caption{Conductance change with magnetic field for Sn$_{0.55}$In$_{0.45}$Te at 5~K. Circles denote experimental data; line is a fit to the WAL formula (see text) with $\alpha = 0.82$ and $l_\phi = 80$~nm. Inset shows the temperature dependence of $\alpha$; dashed line shows an extrapolation to low temperature.
\label{fg:cond}}
\end{center}
\end{figure}

In Fig.~\ref{fg:cond} we plot the experimental $\Delta G$ obtained at 5~K.  The line through the data points is a fit with Eq.~(\ref{eq:G}), which yields the parameters $\alpha = 0.82$ and $l_\phi=80$~nm.  The value of $\alpha$ is temperature dependent, as shown in the inset; it extrapolates towards $\sim2$ at low temperature.

\section{Conclusion}

We have used transport measurements to study the normal state of Sn$_{1-x}$In$_x$Te crystals across much of the composition range for which superconductivity occurs.  We have confirmed that the dominant carrier type changes from hole-like to electron-like near $x\sim0.25$.  The observations of quantum oscillations and a bulk WAL response in the magnetoresistance at $x=0.45$ provide evidence for the hole-like states that have been detected by ARPES about the $L$ points of the bulk Brillouin zone.  Hence, hole-like and electron-like carriers coexist and all contribute to the transport.

In modeling the doping dependence of the Hall effect, we considered a picture in which the In $5s$ states sit somewhat below the chemical potential of SnTe.  At low concentration, these states behave as if they are localized, so that the chemical potential moves lower in the valence band.  With increasing concentration, the In $5s$ levels begin to hybridize with one another, and these electron-like states gain some mobility.   In the future, it would be interesting to see this picture tested with spectroscopic measurements, with a particular focus on characterizing the electron-like states.

This mixture of carriers is of interest with respect to the nature of the superconductivity.  The superconducting transition temperature rises continuously with In concentration across the crossover in dominant carrier type \cite{eric09,bala13,zhon13,hald16}, so presumably both kinds of carriers can contribute to the condensate.   Is the presence of multiple bands relevant to the pairing mechanism? Or, given the modest carrier mobility, are the interactions with the lattice of a more local character?  It was noted quite some time ago that the non-ionic bonding character of IV-VI compounds with the rock salt structure leads to a significant electron-phonon interaction \cite{luco73,litt79}.  Indeed, an enhanced damping has been observed for low-energy transverse acoustic phonons in Sn$_{0.8}$In$_{0.2}$Te \cite{xu15}.    What role does the strong electronic polarizability play in the localization/delocalization of the In $5s$ states, and how does this relate to evidence for strong-coupling superconductivity \cite{hald16}?  Of course, there is also the question of whether there is any topological character to the superconducting state \cite{sasa12,smyl18}. There is clearly more to explore in this system.

\section{acknowledgments}

{\newr We thank Emil Bo\v{z}in for helpful input.}  Work at Brookhaven is supported by the Office of Basic Energy Sciences, Division of Materials Sciences and Engineering, U.S. Department of Energy under Contract No.\ {DE-SC}0012704. R.D.Z., G.G., and J.M.T. were supported by the Center for Emergent Superconductivity, an Energy Frontier Research Center.

\bibliography{topological}

\end{document}